\documentclass[a4,12pt]{article}
\usepackage{graphicx}
\newcommand{\abs}[1]{\left| #1 \right|}

\newcommand{\nn}{\nonumber\\}
\newcommand{\ket}[1]{\left| #1 \right>}
\newcommand{\kakko}[1]{\left( #1 \right)}
\newcommand{\ckakko}[1]{\left\{ #1 \right\}}

\newcommand{\dz}[1]{\frac{d #1}{2\pi i}}
\newcommand{\fron}[1]{\frac{1}{#1}}
\newcommand{\Q}{Q_{\rm B}}
\newcommand{\rL}{{\rm L}}

\newcommand{\CL}{{C_\rL}}

\newcommand{\JB}{J_{\rm B}}

\newcommand{\Cl}{{{\rm C}_{\rm left}}}

\renewcommand{\thepage}{}
\makeatletter
\@addtoreset{equation}{section}
\renewcommand{\theequation}{\thesection.\@arabic\c@equation}
\makeatother
\renewcommand{\thefootnote}{\fnsymbol{footnote}}
\newcommand{\bra}[1]{\left<#1 \right|}

\newcommand{\half}{\frac{1}{2}}

\newcommand{\Ut}{U(t)^{-1}}
\newcommand{\bh}{\bra{h}}
\newcommand{\kh}{\ket{h}}

\begin{document}
\begin{titlepage}
\title{
\vspace*{-4ex}
\hfill{\normalsize hep-th/0506083}\\
\vspace{4ex}
\bf Exploring Vacuum Manifold\\ of Open String Field Theory
\vspace{5ex}}
\author{Yuji {\sc Igarashi}$^{1,}$\footnote{E-mail address:
igarashi@ed.niigata-u.ac.jp},\ \ 
Katsumi {\sc Itoh}$^{1,}$\footnote{E-mail address:
itoh@ed.niigata-u.ac.jp},\ \ 
Fumie {\sc Katsumata}$^{2,}$\footnote{E-mail address:
katsumata@asuka.phys.nara-wu.ac.jp}
\vspace{1ex}\\
Tomohiko {\sc Takahashi}$^{2,}$\footnote{E-mail address:
tomo@asuka.phys.nara-wu.ac.jp}\ \ and 
Syoji {\sc Zeze}$^{3,}$\footnote{E-mail address:
zeze@yukawa.kyoto-u.ac.jp}
\vspace{4ex}\\
$^1${\it Faculty of Education, Niigata University, Japan}\\
$^2${\it Department of Physics, Nara Women's University, Japan}\\
$^3${\it Yukawa Institute for Theoretical Physics, Kyoto University, Japan}}

\maketitle

\begin{abstract}
The global symmetry generated by $K_n$ is a subgroup of the stringy
 gauge symmetry.  We explore the part of the vacuum manifold related by
 this symmetry.  A strong evidence is presented that the analytic
 classical solutions to the cubic string field theory found earlier in
 Refs.~\cite{rf:TT, rf:KT} are actually related by the symmetry and,
 therefore, all of them describe the same tachyon vacuum.  Some
 remaining subtlety is pointed out.
\end{abstract}
\end{titlepage}

\newpage
\renewcommand{\thepage}{\arabic{page}}
\renewcommand{\thefootnote}{\arabic{footnote}}
\setcounter{page}{1}
\setcounter{footnote}{0}
\baselineskip=19pt plus 0.2pt minus 0.1pt
\section{Introduction}

String field theory (SFT) is expected to provide a framework where
distinct string backgrounds can be studied in terms of some universal
set of underlying degrees of freedom.  For example, it was conjectured
that SFT admits the closed string vacuum solution describing the state
after all unstable D-branes are allowed to decay via
condensation of the tachyon \cite{rf:SenUniv,rf:Sen}.
If it happens, no open string excitations
appear around the tachyon vacuum, and the energy difference between the
originally defined D-brane vacuum and the tachyon vacuum is precisely
equal to the brane tension \cite{rf:SZ-tachyon,rf:Senreview}.

It is highly desirable to find analytic candidates for the classical
solution describing the tachyon vacuum. They will make it possible not
only to justify the above conjecture but to give valuable insights into
the structure of the vacuum state in SFT. Such a candidate was discussed
in Refs.~\cite{rf:TT, rf:KT}. There, given was a series of one
parameter families of classical solutions associated with the functions
$h_{a}^{l}(w)$.  The solutions were constructed with these functions
which specify combinations of the BRS current and the ghost field
operated on the identity state.  They are labeled by an integer $l$, and
classified in such a way that non-trivial solutions emerge for the
function $h_{a_{b}}^{l}(w)$~with the boundary value $a=a_{b}$, while all
other solutions associated with $h_{a}^{l}(w)~(a \neq a_{b})$~become
gauge transformations of the D-brane vacuum. In addition to these, we
have constructed another class of solutions with higher order
zeros~\cite{rf:Igarashi4th}.

Based on the analysis on the cohomology \cite{rf:KT,rf:Igarashi4th},
scattering amplitudes \cite{rf:TZ1} and the potential
height \cite{rf:Igarashi4th,rf:tomo}
around each solution, we understand that all of the above 
non-trivial solutions are equally valid candidates for the tachyon
vacuum.  This makes us to believe that all of them are actually
equivalent and related by appropriate symmetry transformations present
in the SFT action.  To clarify this point is the main motivation of this
paper.

The SFT action has the gauge symmetry \cite{rf:CSFT} and it is also
invariant under transformations generated by particular combinations of
the Virasoro operators, $K_n = L_{n} - (-)^{n}L_{-n}$
\cite{rf:SCSFT,rf:LPP1,rf:LPP2}.  Later we will see that the symmetry
generated by $K_n$
belongs to the ``global'' part of the gauge symmetry.  The solutions, if
they are really equivalent, are to be related by some gauge
transformations.  In this paper, we find operators that transform the
higher $l$ solutions down to the $l=1$ solution.  This is the main
result of this paper.  The operators are written in terms of generators
$K_n$ with even $n$. One may wonder what would happen to the inverse of
the relations.  When we go back to a higher $l$ solution from a lower
$l$ solution, we encounter a subtlety, that will be explained in a later
section and further discussion will be found in the last section.

This paper is organized as follows.  In the next section, we briefly
review how analytic classical solutions to the SFT action are
constructed in Refs.~\cite{rf:TT, rf:KT}.  The action expanded around a
solution carries a new BRS charge in its kinetic term.  The charge seems
to have a peculiar ghost structure.  However, it will be explained that
this is naturally understood from the first quantization point of view.
In section 3, we discuss the symmetries of the SFT action.  In
particular, the deformation of a classical solution under
$K_n$-transformations are described.  Since the interaction vertex is
invariant under the transformation, the change of classical solution has
an effect only on the kinetic term, or the BRS charge, in the expanded
action.  Therefore, when we have actions expanded around two classical
solutions that are related by this type of symmetry, we ought to be able
to find an appropriate transformation between two BRS charges.  This is
the subject of the section 4.  We construct the operator $U$ relating
the BRS charges for the $l>1$ and $l=1$ classical solutions and some
properties of the operator are reported.  The operator has a well-defined normal
ordered expression and generate a sound string field
transformation on the component fields.
The last section is devoted to discussions.  We have added four
appendices.  The appendix A is to evaluate the action for the string
field configuration in a pure gauge form. Some technical points in
relation to section 3 are explained in appendix B.  In appendix
C, matrix elements of $U$ are calculated.  The universal solutions due
to \cite{rf:TT, rf:KT} are not in the Siegel gauge as shown in appendix
D.

\section{Classical Solutions in String Field Theory}

This section is a short summary of how analytic classical solutions are
constructed and what makes us believe that they really correspond to the
tachyon vacuum.  Though the construction of solutions for SFT looks quite
non-trivial, the first quantization point of view provides us
with an intuitive picture of the solutions: it also helps us to
understand the constraint embedded in the new BRS charge, defined on a
non-trivial solution.  The first quantization point of view will be
explained in the second subsection.

\subsection{Classical solutions and gauge transformations}

We summarize our construction of the classical
solutions, paying special attention to gauge transformations in cubic
string field theory (CSFT). 

The action in the original CSFT \cite{rf:CSFT},\footnote{For later
convenience, we write the BRS charge dependence of the action as $S[\Psi,Q_B]$.}
\begin{eqnarray}
 S[\Psi,Q_B]=-\frac{1}{g^2}\int \left(\frac{1}{2}\Psi*\Q\Psi
+\frac{1}{3}\Psi*\Psi*\Psi\right), 
\label{Eq:action}
\end{eqnarray}
is characterized by the BRS charge $\Q$ and the $*$-product, which enjoy
the following properties:
\begin{enumerate}
\item $\int Q_B A =0$;
\item $Q_B (A*B) = (Q_B A)*B + (-)^{|A|} A* Q_B B$;
\item $(A * B) * C = A * ( B * C )$;
\item $\int A * B = (-)^{|A| |B|} \int B * A$.
\label{Eq: algebraic structure}
\end{enumerate}
By $|A|$, we denote the Grassmannian even-oddness of the string field
$A$: $|A|=+1~(-1)$, when A is Grassmann even (odd).  

The action (\ref{Eq:action}) is an analogue to the integration of the
Chern-Simons three form \cite{rf:CSFT,rf:SCSFT}: the string field
$\Psi$, the integration $\int$ and the BRS charge $\Q$ correspond to the
connection $A$, the integration of differential forms $\int$ and the
exterior derivative $d$ of ordinary differential geometry.  Similarly to
the Chern-Simons action, it is easy to show that the action
(\ref{Eq:action}) transforms as
\begin{eqnarray}
 S[\Psi', Q_B]=S[\Psi, Q_B]+S[g^{-1}*\Q g, Q_B]
\label{Eq:daction}
\end{eqnarray}
under the gauge transformation
\begin{eqnarray}
 \Psi'=g^{-1}*\Q g+g^{-1}*\Psi*g.
\label{stringy gauge tf}
\end{eqnarray}
Here, the string functional $g$ is an element of the stringy gauge group
in which the multiplication law is given by the star product. This
stringy gauge group is expected to have much richer structure than that
of the ordinary Yang-Mills theory.  Eq.~(\ref{Eq:daction}) reminds us of
a similar expression for the Chern-Simon theory (based on a compact
group).  Though it is tempting to think of a concept of homotopy class
for this stringy gauge group, we do not have much more to discuss along
this direction.  For the purpose of the present paper, it is suffice to
know that the second term in eq.~(\ref{Eq:daction}) vanishes for the
functional $g$ connected to the identity $I$ via a continuous
deformation (see appendix A for the proof).  In other words, the action
is invariant under such a gauge transformation.

The equation of motion is given by the variational principle,
\begin{eqnarray}
 \Q\Psi+\Psi*\Psi=0.
\label{Eq:eqmotion}
\end{eqnarray}
The lhs of (\ref{Eq:eqmotion}) is in the form of the ``field strength''.
Therefore, at least formally, ``flat connections'' are classical
solutions to string field theory.  Let us write such a classical
solution as
\begin{eqnarray}
 \Psi_0=g_0^{-1}*\Q g_0,
\label{Eq:classical solution}
\end{eqnarray}
where $g_0$ is a (group-valued) string functional.

If we expand the string field around the classical solution
(\ref{Eq:classical solution}) as
\begin{eqnarray}
 \Psi= g_0^{-1}*\Q g_0 +\Phi,
\end{eqnarray}
the action (\ref{Eq:action}) becomes
\begin{eqnarray}
 S[\Psi, Q_B]=S[g_0^{-1}*\Q g_0, Q_B]
-\frac{1}{g^2}\int \left(\frac{1}{2}\Phi*\Q' \Phi
+\frac{1}{3}\Phi*\Phi*\Phi\right),
\end{eqnarray}
where the new BRS charge $\Q'$ is defined as
\begin{eqnarray}
 \Q' A \equiv \Q A +g_0^{-1}*\Q g_0*A-(-1)^{\abs{A}} A*g_0^{-1}*\Q g_0,
\end{eqnarray}
for an arbitrary string field $A$.

After having considered the general structure of the string field theory
action, let us explicitly describe the classical solutions given in
Refs.\cite{rf:TT, rf:KT}.

Consider the specific gauge functional \cite{rf:TT,rf:KT,rf:TZ1}
\begin{eqnarray}
 g_0(h)&=&\exp (-q_{\rm L}(h)I)\nn
&=& I-q_{\rm L}(h)I+\frac{1}{2!}\,
q_{\rm L}(h)I*q_{\rm L}(h)I+\cdots,
\label{Eq:gauge-functional}
\end{eqnarray}
where the operator $q_{\rm L}$ is defined in terms of the ghost number
current $J_{\rm gh}(w)$ and a function $h(w)$ satisfying
$h(\pm i)=0$ and $h(-1/w)=h(w)$:
\begin{eqnarray}
 q_{\rm L}(h)=\int_{C_{\rm left}}\,\frac{dw}{2\pi i}
h(w)\,J_{\rm gh}(w).
\end{eqnarray}
The integration path indicated by the subscript $C_{\rm left}$ is over
the left half of the string, ie, $-\pi /2 < \sigma < \pi /2$ on the unit
circle for the variable, $w=e^{i \sigma}$.
The gauge functional (\ref{Eq:gauge-functional}) gives rise to a
classical solution $\Psi_0(h)= g_0(h)^{-1}*\Q\,g_0(h)$, which may be rewritten
as 
\begin{eqnarray}
\label{Eq:sol}
\Psi_0(h)= Q_{\rm L}(e^h-1)I-C_{\rm L}((\partial\, h)^2\,e^h)I,
\label{Eq:TT-solution}
\end{eqnarray}
where the operators $Q_{\rm L}$ and $C_{\rm L}$ are defined with the
BRS current and the ghost field:
\begin{eqnarray}
 Q_{\rm L}(f)=\int_{C_{\rm left}}\frac{dw}{2\pi i}\,
f(w)\,J_{\rm B}(w),\ \ \ 
 C_{\rm L}(f)=\int_{C_{\rm left}}\frac{dw}{2\pi i}\,
f(w)\,c(w).
\label{Eq:QL and CL}
\end{eqnarray}
The solution (\ref{Eq:TT-solution}) has a well-defined Fock space
expression in the universal subspace, spanned by the matter Virasoro
generators and ghost oscillators acting on the $SL(2,R)$ invariant
vacuum.  Therefore, it would be appropriate to call the solution
(\ref{Eq:TT-solution}) as the universal solution.

For the solution (\ref{Eq:TT-solution}), the new BRS charge
can be expressed as 
\begin{eqnarray}
\label{Eq:newBRS}
\Q'=Q(e^h)-C((\partial\,h)^2\,e^h).
\end{eqnarray}
The operators $Q$ and $C$ are defined as
\begin{eqnarray}
  Q(f)=\oint \frac{dw}{2\pi i}\,
f(w)\,J_{\rm B}(w),\ \ \ 
 C(f)=\oint \frac{dw}{2\pi i}\,
f(w)\,c(w).
\end{eqnarray}
Here the integrations are over the unit circle.

Formally, the solution (\ref{Eq:sol}) is in a pure gauge form, and,
therefore, could be gauged away. As shown in \cite{rf:TT}, however, this
is not always the case: some non-trivial solutions emerge at the
boundary of one parameter deformation of a certain class of functions
chosen for $h$.  The functions are given by \cite{rf:TT,rf:KT}
\begin{eqnarray}
\label{Eq:hma} 
h_a^l(w)=\log\left\{
1-\frac{a}{2}(-1)^l\left(
w^l-\left(-\frac{1}{w}\right)^l\,\right)^2\right\}\ \ \ (l=1,2,3,\cdots).
\end{eqnarray}
We have a series of functions labeled by the integer $l$.  Accordingly,
we have a series of non-trivial solutions associated with the
functions.  The solutions will be addressed as TTK solutions in this paper.

Now we describe evidences suggesting that the TTK solutions are really
non-trivial.  The reality of the function, for $w$ on the unit circle,
restricts the parameter $a$ to be $a\geq -1/2$.  This condition
guarantees the hermiticity of the corresponding classical solution as
well as the new BRS charge, as expected from eq.~(\ref{Eq:newBRS}).
Since $h^l_{a=0}(w)=0$ and the corresponding classical solution
vanishes, it would be reasonable to expect that the solution written
with $h_a^l$ is a trivial pure gauge for sufficiently small $a$.
Indeed, we know the following facts for $a>-1/2$:
\begin{enumerate}
 \item The expanded action can be transformed back to the action
       (\ref{Eq:action}) \cite{rf:TT};
 \item The new BRS charge provides us with the same cohomology as the
       original BRS charge \cite{rf:TT,rf:KT,rf:Igarashi4th};
 \item The expanded theory reproduces ordinary open string scattering
       amplitudes \cite{rf:TZ1};
 \item A numerical study with the level truncation technique shows that
       the expanded theory has a non-perturbative vacuum and its vacuum
       energy tends to the D-brane tension as the truncation level
       increases \cite{rf:tomo}.
\end{enumerate}
These facts are consistent with our expectation that the solution
is trivial pure gauge.
However, at the boundary $a=-1/2$, the expanded theory shows
completely different properties\footnote{It is not known whether the
non-trivial solutions with $a=-1/2$ can still be written in the form
$g^{-1}*\Q g$.  If it is the case, we may consider that the
solutions are obtained by ''large gauge transformations.''  That would
be a strong evidence for some topological structure of the stringy gauge
group.}:
\begin{enumerate}
 \item[5.] The new BRS charge has the vanishing cohomology in the Hilbert space
       with the ghost number one \cite{rf:KT,rf:Igarashi4th};
 \item[6.] The vanishing of open string scattering amplitudes, the
	   result is consistent with the absence of open string
	   excitations ({\it no open string theorem}) \cite{rf:TZ1};
 \item[7.] We can show numerically that the non-perturbative vacuum
	   found for $a>-1/2$ disappears as the parameter $a$ approaches
	   to $-1/2$ \cite{rf:tomo}.
\end{enumerate}
The above results implies that the solution with $a=-1/2$ indeed
corresponds to the tachyon vacuum.

\subsection{Interpretation of new BRS charges in the first quantized theory}

The emergence of the non-trivial theory for $a=-1/2$ can also be seen from
the first quantization point of view.

In the
original action (\ref{Eq:action}), we have the Kato-Ogawa's BRS charge
\begin{eqnarray}
\Q = \oint \frac{dw}{2\pi i}\, 
\left[c(w) T_{X}(w) + (bc~\partial c)(w) \right],
\end{eqnarray}
where $T_{X}(w)$ is the stress tensor for the string coordinates $X$ and
$b$ is the antighost. The BRS charge $\Q$ is known to be constructed
directly from the first-class constraint $T_{X}(w) \approx 0$ \cite{rf:HT}. 

We now consider a modification of the constraint surface by multiplying
a function, $e^{h(w)}$. Then, the modified BRS charge constructed from
the constraint $e^{h(w)}~T_{X}(w) \approx 0$ takes of the form
\begin{eqnarray}
\Q' = \oint \frac{dw}{2\pi i}\, e^{h(w)}
\left[c(w) T_{X}(w) + (bc~\partial c)(w) + \frac{1}{2}c(w)\{(\partial~h)^{2} + 3 ( \partial^{2}~h)\}
\right].
\label{Eq:modBRS}
\end{eqnarray}
Here the term linear in the ghost is needed to ensure the nilpotency
condition $(\Q')^2=0$. The expression (\ref{Eq:modBRS}) coincides with
(\ref{Eq:newBRS}) where the BRS current is given by
$j_{B}(w)=c(w)[T_{X}(w)+ (b\partial c)(w) +3/2 \partial^2 c(w)]$.  It
means that the replacement of the BRS charge $\Q$ by $\Q'$ corresponds
to the replacement of the constraint $T_{X}(w) \approx 0$ by
$e^{h(w)}~T_{X}(w) \approx 0$.  This change of the constraint can be
absorbed by a redefinition of ghost and antighost:
\begin{eqnarray}
c(w) \rightarrow c(w)e^{h(w)} &=& e^{q(h)} c(w) e^{-q(h)}, \nonumber\\
b(w) \rightarrow b(w)e^{-h(w)} &=& e^{q(h)} b(w) e^{-q(h)},
\end{eqnarray}
so that 
\begin{eqnarray}
e^{q(h)} \Q e^{-q(h)} = \Q'.
\label{Eq:QBrelation}
\end{eqnarray}
This relation holds, of course, only if the operator $e^{q(h)}$ with 
\begin{eqnarray}
 q(h)=\oint \frac{dw}{2\pi i}\,
h(w)\,J_{\rm gh}(w)
\end{eqnarray}
is well-defined.  The operator $e^{q(h^{l}_{a})}$, with the function
given in (\ref{Eq:hma}), is well-defined for $a>-1/2$, but not for
$a=-1/2$ \cite{rf:TT, rf:KT}.  Whether the similarity transformation
(\ref{Eq:QBrelation}) makes sense or not depends on the distribution of
zeros \cite{rf:Drkr2,rf:zeze} of the function $\exp (h^{l}_{a}(w))$:
all zeros are distributed off the the unit circle $|w|=1$ for $a>-1/2$,
while they merge on the unit circle for $a=-1/2$. This change in the
distribution of zeros may be related to the non-trivial modification of the
constraint in first quantized theory: the constraint surface given by
$\exp (h^{l}_{-1/2}(w))~T_{X}(w) \approx 0$ becomes physically distinct
from the original surface $T_{X}(w) \approx \exp
(h^{l}_{a>-1/2}(w))~T_{X}(w) \approx 0$.\\

In this section, the properties of our classical solutions are
summarized.  As we have seen, there present an infinite number of
non-trivial solutions and various results suggest that they all describe
the tachyon vacuum.  If it is really the case, they are equivalent with
each other and related via the symmetries of CSFT.  The symmetries of
CSFT are the subject of the next section.

\section{Symmetries of CSFT and Classical Solutions}

The CSFT action has a subalgebra of the Virasoro algebra as its symmetry
\cite{rf:SCSFT,rf:LPP1,rf:LPP2}.  Here we will see that this symmetry
may be considered 
as a subgroup of the ``global'' part of the stringy gauge symmetry.
Therefore it provides a way to relate classical solutions and, thereby
the SFT actions defined on them.

The subalgebra is generated by $K_{n}=L_{n}- (-)^{n}L_{-n}$, where
$L_{n}$ is the Virasoro operator. Using the properties
\cite{rf:SCSFT,rf:LPP1,rf:LPP2}, 
\begin{eqnarray}
\int K_n A &=& 0~~~{\rm for~all}~A, \nonumber\\
K_n (A * B) &=& (K_n A) * B + A * K_n B \label{Kn property 2},
\end{eqnarray}
it is easy to show the invariance of the action under an infinitesimal 
transformation generated by $K_{n}$ :
\begin{equation}
\delta S \propto \int K_n \Bigl( \frac{1}{2} \Psi * Q_B \Psi 
+ \frac{1}{3} \Psi * \Psi * \Psi \Bigr)=0.
\end{equation}

We now consider a particular type of gauge transformation and find it to
be a finite form of $K_n$-transformation written as
\begin{equation}
\Psi^{\prime} \equiv e^{K(v)} \Psi
\label{finite tf by K_n},
\end{equation}
with $K(v) \equiv \sum_{n>0} v_n K_n$.  The parameters $v_n$ will be
specified below.

Let us take $u_L(f)\equiv \exp\Bigl(-{\cal T}_L(f) I \Bigr)$ as a gauge
functional. The operator ${\cal T}_L(f)$ is defined as
\begin{equation}
{\cal T}_{\rm L}(f)=\int_{C_{\rm left}}\frac{dw}{2\pi i}\,
f(w)\,T(w),
\end{equation}
similarly to eq.~(\ref{Eq:QL and CL}), with the total energy-momentum
tensor $T(w)$.  It is easy to see that the gauge transformation
(\ref{stringy gauge tf}) with $u_L(f)$ can be written as
\begin{eqnarray}
\Psi' &=& u_L^{-1} * \Psi * u_L = U(f) \Psi, 
\label{global gauge tf}\\
&{}&U(f) \equiv \exp \Bigl( \oint \frac{dw}{2\pi i}\,f(w)\,T(w) \Bigr).
\label{U(f) by T(w)}
\end{eqnarray}
Note that the first term in (\ref{stringy gauge tf}) with the BRS charge
is absent in the above expression since $[Q_B,~{\cal T}_L(f)]=0$ and
$Q_B I =0$.  In deriving the last expression of eq.~(\ref{global gauge
tf}), we used the properties of the half splitting operators in
eqs.~(\ref{Tl Tr}) and (\ref{Eq:TABTI}) that require the function $f(w)$
to satisfy the condition, $f(w)=(dw/d {\tilde w}) f({\tilde w})$ for
${\tilde w}= - 1/w$.\footnote{Some more detailed derivation of
(\ref{global gauge tf}) is described in appendix B.}  Expanding the
function as $f(w)= \sum_n v_n w^{n+1}$, we find the relation,
$v_n=v_{-n}(-)^{n+1}$, from the condition.  Using this relation, it
is easy to see that the integral in eq.~(\ref{U(f) by T(w)}) becomes the
operator $K(v) \equiv \sum_{n>0} v_n K_n$.

The absence of the term $u_L(f)*Q_B u_L(f)^{-1}$ in the expression
(\ref{global gauge tf}) allows us an interesting interpretation of the
transformation (\ref{finite tf by K_n}).  The BRS charge $Q_B$
corresponds to the external derivative $d$ in the Chern-Simons theory.
A gauge transformation in the CS-theory with the absence of the
derivative term is simply a global transformation.  Similarly, a stringy
gauge transformation without the first term in (\ref{stringy gauge tf})
may be considered to be a ``stringy global transformation'' in SFT.
Note that, strictly speaking, a global transformation is not necessarily
a finite $K_n$-transformation.  The type of global transformations
written as (\ref{finite tf by K_n}) forms
a specific subgroup in the stringy global transformations.  In the rest
of the paper, we discuss only this subgroup of the global
symmetry.

Now, we apply the global transformations discussed above on a given
classical solution.
Let us find the action for the fluctuation $\Phi$ around a classical
solution $\Psi_0$.  Substituting $\Psi \equiv \Psi_0+\Phi$ into the action
$S[\Psi,~Q_B]$, we obtain
\begin{eqnarray}
S[\Psi_0+\Phi, Q_B]&=&S[\Psi_0, Q_B]+S[\Phi, Q_B(\Psi_0)].
\label{expand around the sol}
\end{eqnarray}
Here $Q_B(\Psi_0)$ is defined as 
\begin{eqnarray}
Q_B(\Psi_0) A &\equiv& Q_B A + \Psi_0 * A - (-)^{|A|}A*\Psi_0
\label{BRS charge on Phi}
\end{eqnarray}
on a string field $A$.  The nilpotency follows from the equation of motion for $\Psi_0$.

Once we find a classical solution, we can, at least formally, obtain
other solutions related by the gauge symmetry.  As for solutions related
as eq.~(\ref{finite tf by K_n}), it is easy to see the following
statements to hold:
\begin{enumerate}
 \item If $\Psi_0$ solves the string equation of motion, ie, $Q_B \Psi_0
       + \Psi_0 * \Psi_0=0$, then $\Psi'_0 \equiv e^{K(v)}\Psi_0$ is
       also a solution;
 \item Furthermore the BRS charges defined around two solutions are
       related as
\begin{equation}
 e^{-K(v)}~Q_B(e^{K(v)}\Psi_0)~e^{K(v)}=Q_B(\Psi_0).
\label{relating two BRS charges}
\end{equation}
\end{enumerate}
Since the transformation (\ref{finite tf by K_n}) leaves the action
invariant, the first statement is trivial.  Technically speaking, it
can be shown by using the property,  
\begin{equation}
(e^{K(v)}A) * (e^{K(v)}B) = e^{K(v)}(A * B),
\label{K(v) property}
\end{equation}
which is obtained from eq.~(\ref{Kn property 2}).  The second statement follows
from the relation,
\begin{eqnarray}
Q_B(\Psi'_0) e^{K(v)} {A} = e^{K(v)} Q_B(\Psi_0){A},
\label{BRS charges relation}
\end{eqnarray}
on a generic string field $A$.  Eq.~(\ref{BRS charges relation}) may be
derived from the definition of the BRS charge given in eq.~(\ref{BRS charge on
Phi}).

\vspace{5mm}

Now we would like to see how a universal solution may be deformed by the
action of the $K_n$ operators.  Leaving discussions of the finite
transformation in the next section, here we consider the change of a
solution under an infinitesimal transformation and discuss its
implication.  

The generic form of the classical solutions obtained in
Refs. \cite{rf:TT,rf:KT} is given as
\begin{eqnarray}
 |\Psi_0 \rangle &=& Q_L(F)|I\rangle + C_L(G)|I\rangle
\label{generic form of classical solution},\\
 &{}&G(w)=-\frac{(\partial F(w))^2}{1+F(w)},
\label{rf:F and G}
\end{eqnarray}
where $F(z)$ is an analytic function satisfying the relations
$F(-1/w)=F(w)$ and $F(\pm i)=0$.

It is easy to see that the action of $e^{K(\varepsilon)}$ on the
universal solution $|\Psi_0 \rangle$ produces yet another universal
solution, at least in the first order in deformation parameters
$\varepsilon_n$.  The effect of the operator appears as a change in the
function $F(z)$,
\begin{eqnarray}
 |{\tilde \Psi_0} \rangle = e^{K(\varepsilon)}|\Psi_0 \rangle 
&\sim& Q_L({\tilde F})|I\rangle + C_L({\tilde G})|I\rangle,
\label{infinitesimal K-tf}\\
{\tilde F}(w) &\equiv& F(w)-\sum_{n=1}^{\infty}\varepsilon_n u_n(w)\partial
 F(w),
\label{change in F}\\
&{}&u_n(w)\equiv w (w^n-(-)^nw^{-n}).
\end{eqnarray}
It is easily confirmed that the deformed function also satisfy two
conditions for a classical solution.\footnote{In eq.~(\ref{change in
F}), we observe that no choice for a set of parameters leaves the
function invariant.  This implies that the symmetry generated by $K_n$
does not survive, at least, in its original form.}

The above calculation shows that we may explore the submanifold of
classical solutions by the action of $e^{K(v)}$.  This opens a
possibility to relate different solutions written in the same form as
described in (\ref{generic form of classical solution}).  In particular,
the series of solution constructed in \cite{rf:TT,rf:KT} may be related
by the operator $e^{K(v)}$ with its parameters appropriately chosen.  Of 
course, this cannot be realized by infinitesimal transformations
discussed here and we have to consider finite transformations.  

We make another observation that supports this idea.
Consider the SFT defined around the classical solution
$\Psi_{0}(h^{l}_{a})$ written as eq.~(\ref{Eq:sol}) with the function in
(\ref{Eq:hma}).  After taking the limit of $a \rightarrow -\frac{1}{2}$,
we have the action $S[\Phi, Q_B^{(l)}]$ with the BRS charge given as
\begin{eqnarray}
Q_B^{(l)} &=& \frac{1}{2}Q_B + \frac{(-)^l}{4}(Q_{2l}+Q_{-2l}) 
+ 2l^2\Bigl(c_0 -\frac{(-)^l}{2}(c_{2l} +c_{-2l})\Bigr)\nonumber\\
&=& Q ( F^{(l)} ) + C( G^{(l)} ),
\label{lth BRS charge}
\end{eqnarray}
where the moments of the BRS current are defined in the expansion,
$J_{B}(w) \equiv \sum_{n} Q_{n}w^{-n-1}$, and the functions $F^{(l)}(w)$
and $G^{(l)}(w)$ are given as
\begin{eqnarray}
F^{(l)}(w) &\equiv& \frac{(-)^l}{4} \Bigl(w^l+(-w)^{-l} \Bigr)^2,~~~
G^{(l)}(w) \equiv -l^2w^{-2}(-)^l  \Bigl(w^l-(-w)^{-l} \Bigr)^2.
\end{eqnarray}

If the classical solutions labeled by $l$ and $m~(l \ne m)$ are really
related by the finite form of the $K_n$ symmetry, the BRS charges in the
actions $S[\Phi, Q_B^l]$ and $S[\Phi, Q_B^m]$ are to be related as
\begin{eqnarray}
Q_B^{(m)}= e^{-K(v)} \Q^{(l)} e^{K(v)},
\label{Eq:Similarity tf of BRS charge}
\end{eqnarray}
with some operator $e^{K(v)}$, as we stated in (\ref{relating two BRS
charges}).  From eq.~(\ref{lth BRS charge}) and the commutation
relation, $[K_{n},~Q_{2l}] = -2l (Q_{2l+n} - (-)^{n} Q_{2l-n})$, we
realize that $K_{n}$ with $n=$ even are to be used for the purpose.  So
it is possible for eq.~(\ref{Eq:Similarity tf of BRS charge}) to hold.
We are to find out whether we may choose proper parameters so
that eq.~(\ref{Eq:Similarity tf of BRS charge}) holds.  Further
discussion of the transformation (\ref{Eq:Similarity tf of BRS charge})
will be given in the next section.

Before closing this section, we introduce operators which play an
important role in relating BRS charges:
\begin{equation}
U_{2l}(t) =\exp \Bigl[-\frac{(-)^l}{4l}{\rm
ln}\Bigl(\frac{1-t}{1+t}\Bigr) K_{2l} \Bigr]~~~~~~~~(l=1, 2, ...).
\label{U2l op}
\end{equation}
When ordered with respect to the Virasoro operators, they are expressed
as
\begin{equation}
U_{2l}(t)= \exp\Bigl({\frac{-(-)^lt}{2l}\cdot L_{-2l}}\Bigr)
\exp\Bigl(  {\frac{1}{2l} {\rm ln}(1-t^2) \cdot L_{0}}  \Bigr)
\exp\Bigl({\frac{(-)^lt}{2l}\cdot L_{2l}}\Bigr).
\end{equation}
The operators in eq.~(\ref{U2l op}) generate the conformal
transformations \cite{rf:LPP1},
\begin{equation}
f_{2l}(z, t )=\biggl(
\frac{z^{2l}-(-)^l~t}{1-(-)^l~t z^{2l}}
\biggr)^{\frac{1}{2l}}.
\label{conf tf for U2l}
\end{equation}
A comment is in order.  These conformal transformations are used earlier
\cite{rf:TZ1} in connection with the TTK solutions.  The parameter $t$
introduced here corresponds to $Z(a) \equiv (1+a-{\sqrt{1+2a}})/a$ in
the earlier expression.

\section{Relating classical solutions with different values of $l$}

When two classical solutions are related by the global transformation,
expansions around them produce BRS charges satisfying eq.~(\ref{relating
two BRS charges}).  Here we present a way to construct operators
relating the BRS charges for the TTK classical solutions.

In the last section, we have seen that the BRS charges may be related as
eq.~(\ref{Eq:Similarity tf of BRS charge}) with the operator $K(v)$
written in terms of $K_n$ ($n=$even).  Here we construct the operator
$e^{K(v)}$ that relates charges associated with $l=1$ and $l \ne 1$.  In
the next subsection, it will be shown that the operator is realized in
the limit of a one-parameter family of operators, $U(t)~~(-1 \le t \le
0)$:
\begin{eqnarray}
U(t=0)=1,~~~~~Q^{(l=1)}_B = \lim_{t \rightarrow -1} U(t)Q_B^{(l)}U^{-1}(t).
\label{conditions on U(t)}
\end{eqnarray}

Unfortunately, there seems to be a subtlety in this operator: when
trying to obtain the higher $l$ BRS charge from the $l=1$ charge, we
encounter a problem.  This will be explained briefly in the second
subsection.  It is not clear to us at this moment whether this is just a
technical problem or much deeper one.  In the third subsection, the
properties of the operator $U(t)$ are investigated.  In particular, it
will be shown that it has the well-defined normal ordered expression in
terms of the Virasoro generators even in the limit of $t \rightarrow
-1$.

\subsection{Higher $l$ solutions down to $l=1$ solution} 

Our construction of operators that relate BRS charges is based on the
observation to be explained below.
On the $l$-th BRS charge (\ref{lth BRS charge}), we act the
operator introduced in eq.~(\ref{U2l op}) and find the charge
transformed as,
\begin{eqnarray}
&{}&~~~~~U_{2l}(t)Q_B^{(l)} U_{2l}^{-1}(t) = Q(F_1^{(l)})+C(G_1^{(l)}),
\label{U2l transformed QB}\\
F_1^{(l)}(w,t) &\equiv& F^{(l)}(z)\vert_{z=f_{2l}(w,-t)}= \frac{(-)^l}{4} \Bigl(w^l+(-w)^{-l} \Bigr)^2
\frac{(1+t)^{2}}{\left(1+(-)^{l}t~w^{2l}\right)\left(1+(-)^{l}t~w^{-2l}\right)}\nonumber\\
&=& \frac{1+t}{2} +
 \frac{1-t^{2}}{4}\sum_{n=1}^{\infty}(-)^{n-1}~(-)^{ln}t^{n-1}(w^{2ln} +
 w^{-2ln}),
\label{F of U2l on QB}\\
G_1^{(l)}(w,t) &\equiv& G^{(l)}(z)\vert_{z=f_{2l}(w,-t)} \times \Bigl( \frac{d f_{2l}(w,-t)}{d w} \Bigr)^2 \nonumber\\
&=& -(-)^{l} l^{2}w^{-2}\Bigl(w^l-(-w)^{-l} \Bigr)^2 
\frac{(1+t)^{2}(1-t)^{4}}{\left(1+(-)^{l}t~w^{2l}\right)^{3}\left(1+(-)^{l}t~w^{-2l}\right)^{3}}
\nonumber\\
&=& \frac{2l^{2}(1+t+t^{2})}{w^{2}(1-t^2)}
- \frac{l^{2}}{2 w^{2}(1-t^2)}
\sum_{n=1}^{\infty}(-)^{ln}~g_{n}(t)(w^{2ln} + w^{-2ln}).~~~
\label{G of U2l on QB}
\end{eqnarray}
Here $f_{2l}(w,t)$ is given in eq.~(\ref{conf tf for U2l}) and the
coefficients $g_{n}(t)$ are given by
\begin{eqnarray}
g_{n}(t)=(-t)^{n-1}\Bigl[n^{2} + n + 4(n+1)t-2(n^2-2)t^{2}-4(n-1)t^{3}+(n^{2}-n)t^{4}\Bigr]. 
\label{gnt}
\end{eqnarray}
Substituting eqs.~(\ref{F of U2l on QB}) and (\ref{G of U2l on QB})
into the terms on the rhs of eq.~(\ref{U2l transformed QB}), we obtain
the expression 
of $U_{2l}(t)Q_B^{(l)} U_{2l}^{-1}(t)$ as
\begin{eqnarray}
U_{2l}(t)Q_B^{(l)} U_{2l}^{-1}(t) ~=~ \frac{1+t}{2}Q_B&+& \frac{1-t^{2}}{4}\sum_{n=1}^{\infty}
(-1)^{n(l-1)-1}t^{n-1}(Q_{2ln} + Q_{-2ln}) \nonumber\\
+ \frac{2(1+t+t^{2})l^{2}}{1-t^2} c_{0} 
&-& \frac{l^{2}}{2(1-t^2)} 
\sum_{n=1}^{\infty}
(-)^{nl} g_{n}(t)(c_{2nl} + c_{-2ln}).
\label{explicit UQBU}
\end{eqnarray}
On the rhs, we have higher moments of the BRS currents and pure ghost
terms.

Note that, in the limit of $t \rightarrow -1$, all the terms containing
the moments of BRS currents vanish. The remaining pure ghost terms
become divergent.  So some care should be taken when we take this limit at this
stage.  It would be worth pointing out that the pure ghost terms are
indeed those appeared in the vacuum string field theory.  This fact was
first realized in Refs.~\cite{rf:Drkr2,rf:Drkr1} and utilized recently to regularize
the VSFT \cite{rf:DrukkerOkawa}.

Another important observation on eq.~(\ref{explicit UQBU}) is the fact
that the rhs contains the BRS charge itself.  Since all the TTK
solutions have this property, we may relate various BRS charges via the
expression on the rhs of (\ref{explicit UQBU}).  In concrete, we act
$U_2(t)$ for $l=1$ on the rhs of (\ref{explicit UQBU}) and see what
would come out.  Our expectation is that the result is somewhat close to
the BRS charge for the $l=1$ solution.

The action of $U_2(t)$, on the rhs of (\ref{explicit UQBU}), replaces the
argument of $F_1^{(l)}(w,t)$ by $f_2(w,t)$.  After some calculations, we obtain
\begin{eqnarray}
F_2^{(l)}(w, t) &\equiv& F_1^{(l)}(z,t)\vert_{z=f_2(w,t)}
\nonumber\\
&=&-\frac{1}{4}\Bigl(w-1/w \Bigr)^2 \Bigl( \frac{2}{l+1}\Bigr)^2
\frac{1}{\Bigl(  1+ \frac{l-1}{l+1}w^2\Bigr)\Bigl(  1+ \frac{l-1}{l+1}w^{-2}\Bigr)}
+ O\Bigl((1+t)\Bigr).
\label{F''}
\end{eqnarray}
Similarly we obtain the expression for $G_2^{(l)}(w,t)$.  Note that the
singular behavior in the limit of $t \rightarrow -1$, observed in the
rhs expression of eq.~(\ref{explicit UQBU}), is absent in (\ref{F''}).
The action of $U_2(t)$ has canceled the singular behavior.
In the limit both $F_2^{(l)}(w,t)$ and $G_2^{(l)}(w,t)$ are finite.  So
we may take the limit of $t \rightarrow -1$ in the expression for
$U_2(t)^{-1}U_{2l}(t)Q_B^{(l)}U_{2l}(t)^{-1}U_2(t)$ where $U(t) \equiv U_2(t)^{-1}U_{2l}(t)$:
\begin{eqnarray}
&{}& \lim_{t \rightarrow -1} 
U_2(t)^{-1}U_{2l}(t)Q_B^{(l)}U_{2l}(t)^{-1}U_2(t)
\equiv
 Q({\tilde F}_2^{(l)}) + C({\tilde G}_2^{(l)})~,
\label{UQBU}\\
&{}& ~~~{\tilde F}_2^{(l)}(w) \equiv \lim_{t \rightarrow -1}F_2^{(l)}(w, t)
= -\frac{1}{4}\Bigl(w-1/w \Bigr)^2 \Bigl( \frac{2}{l+1}\Bigr)^2
\frac{1}{\Bigl(  1+ \frac{l-1}{l+1}w^2\Bigr)\Bigl(  1+ \frac{l-1}{l+1}w^{-2}\Bigr)}~,
\label{function F(w)}\\
&{}& ~~~{\tilde G}_2^{(l)}(w) \equiv - 
\frac{(\partial {\tilde F}_2^{(l)}(w))^2}{{\tilde F}_2^{(l)}(w)}~.
\nonumber
\end{eqnarray} 
Here we find the function $(w-1/w)^2/4$ in eq.~(\ref{function F(w)}),
the function for the $l=1$ BRS charge.  The remaining factor appeared in
eq.~(\ref{function F(w)}) may be removed by an appropriate
transformation.  Indeed, we can deform the $l=1$ BRS charge to the form
of eq.~(\ref{function F(w)}):
\begin{eqnarray}
U_2\Bigl(- \frac{l-1}{l+1}\Bigr) Q_B^{(1)}U_2\Bigl(-
 \frac{l-1}{l+1}\Bigr)^{-1}=Q({\tilde F}_2^{(l)})+ C({\tilde G}_2^{(l)})~.
\label{small action of U2}
\end{eqnarray}

Combining eqs.~(\ref{UQBU}) and (\ref{small action of U2}), we finally
reach the $l=1$ BRS charge starting from the higher $l$ charge:
\begin{eqnarray}
U^{-1}_2\Bigl( -\frac{l-1}{l+1}\Bigr) \lim_{t \rightarrow -1} 
\Bigl(
U_2(t)^{-1}U_{2l}(t)Q_B^{(l)}U_{2l}(t)^{-1}U_2(t)
\Bigr) U_2 \Bigl( -\frac{l-1}{l+1}\Bigr)= Q_B^{(1)}~.
\nonumber
\end{eqnarray}
The expression is also rewritten as
\begin{eqnarray}
\lim_{t \rightarrow -1} U(t) Q_B^{(l)} U^{-1}(t) = Q_B^{(1)}~,
\label{final expression}
\end{eqnarray}
where $U(t) \equiv U_2({\tilde t})^{-1}U_{2l}(t)$
with ${\tilde t}$ defined as 
\begin{eqnarray}
{\tilde t} \equiv \frac{-(l-1)+(l+1)t}{(l+1)-(l-1)t}~.
\label{t tilde}
\end{eqnarray}
The parameter in
$U_2$ is now ${\tilde t}$ due to the extra action of
$U_2(-\frac{l-1}{l+1})$ in eq.~(\ref{small action of U2}).

\subsection{$l=1$ solution up to  $l=2$ solution}

A natural question is whether we can obtain the SFT action around the
$l=2$ solution starting from that for $l=1$ solution.  As a relation of
BRS charges, our question may be formulated as follows: is it possible to
find an operator $\cal U$ such that ${\cal U}^{-1}Q_B^{(1)}{\cal
U}=Q_B^{(2)}$?  In this subsection, we explain what we have understood
in relation to this question.

Let us reconsider eq.~(\ref{final expression}) for $l=2$. Before taking the
limit, we write 
\begin{eqnarray}
Q_B^{(1)}(t) \equiv U_2^{-1}({\tilde t}) U_{4}(t)
Q_B^{(2)}U^{-1}_{4}(t)  U_2({\tilde t}) \equiv Q(F_t)+C(G_t) 
\label{Before taking the limit}
\end{eqnarray}
where ${\tilde t}=(3t-1)/(3-t)$ as defined in eq.~(\ref{t tilde}),
and the function $F_t(w)$ is given as
\begin{eqnarray}
F_t(w) = \frac{1}{4}(f^2+f^{-2})^2,~~~ 
f(w; t) \equiv f_4\Bigl( f_2(w; {\tilde t}); -t \Bigr).
\end{eqnarray}
Note here that $F_t(w)$ is slightly different from $F_2^{(l=2)}(w,t)$
since the former now includes the contribution from
$U_2(-\frac{l-1}{l+1})|_{l=2}$ in eq.~(\ref{small action of U2}).  Let
us write the function $F_t(w)$ explicitly,
\begin{eqnarray}
F_t(w) = \frac{1}{4} \cdot \frac{\Bigl(4 {\tilde t} + (1+ {\tilde
t}^2)(w^2+w^{-2})\Bigr)^2}{\displaystyle \left( \frac{9t^2 -14 t
+9}{(t-3)^2}+2{\tilde t}w^2 + \left(\frac{t+1}{t-3}\right)^2 w^4 \right) \Bigl( w
\rightarrow w^{-1} \Bigr) }.
\end{eqnarray}

In the limit of $t \rightarrow -1$, $F_t(w)$ becomes
\begin{eqnarray}
F_t(w) \rightarrow - \frac{1}{4}(w-w^{-1})^2,
\label{limiting Ft}
\end{eqnarray} 
in other words,
\begin{eqnarray}
Q_B^{(2)}(t) \rightarrow Q_B^{(1)},
\nonumber
\end{eqnarray}
the result of the previous subsection.

Now how about the inverse? The relation $U^{-1}(t)Q_B^{(1)}(t)U(t) =
Q_B^{(2)}$ must hold since algebraically $U^{(-1)}(t)U(t)=1$.  Indeed,
we will confirm this relation shortly.  However, the relation is not the
direct answer to our question.  To find an answer, we would rather like
to consider the operator $U^{-1}(t)Q_B^{(1)}(s)U(t)$.  If we could take
the limit of $s \rightarrow -1$ first, then take the other limit, the
operator $\displaystyle \lim_{t \rightarrow -1}U(t)$ would be ${\cal U}$ we are
looking for.

In calculating $U^{-1}(t)Q_B^{(1)}(s)U(t)$, we first consider
$U_2({\tilde t})Q_B^{(1)}(s)U^{-1}_2({\tilde t})$.  The function
$F_s(w)$ in the integrand $Q(F_s)$ is replaced as\footnote{The other
function $G_s(w)$ is also changed accordingly.}
\begin{eqnarray}
F_s(w) \rightarrow F_s(z)\vert_{z=f_2(w;-{\tilde t})}.
\label{U2 action}
\end{eqnarray}
An explicit calculation of the rhs of eq.~(\ref{U2 action}) shows that
the factors in the numerator and denominator of the resultant expression
carries terms with $w^2$ and $w^4$ other than constants.  In the next
step, we let the operator $U_4(t)$ act on the rhs of eq.~(\ref{U2
action}) and the variable $w$ is replaced by $f_4(w;t)$, which has the
fourth order branch cuts.  The factor $w^2$ is replaced by $\Bigl(
f_4(w;t) \Bigr)^2$ and it still has branch cuts.  Generically those
produce the ambiguity in defining the contour integration.  In order to
avoid this difficulty, we require the terms of $w^2$ vanish on the rhs
of eq.~(\ref{U2 action}): this condition is found to be $s=t$.  If we
take $s=t$, there is no ambiguity and we find that
$U^{-1}(t)Q_B^{(1)}(t)U(t)$ is certainly $Q_B^{(2)}$, even before taking
the limit of $t \rightarrow -1$.

We have not been able to make sense of taking the limits of
$U^{-1}(t)Q_B^{(1)}(s)U(t)$ independently.  So we still have not reached
the complete understanding of the relations between various TTK
solutions.

\subsection{Operator $U(t)$ and string field transformation}

Though with some limitation, we have constructed the operator that
relates the $l$-th and the first solutions of TTK solutions in the limit
of $t \rightarrow -1$.  Here we describe some properties of the
operator: the differential equation; some evidence that suggests the
well-definedness of the operator.  For concreteness, we consider the
operator $U(t) \equiv U_2^{-1}({\tilde t})U_4(t)$ relating the first and
the second solutions.  With the operator $U(t=-1)$, we are supposed to
be able to relate the string fields defined around two solutions:
$|\Phi^{(2)}\rangle = U(-1)^{-1}|\Phi^{(1)}\rangle$.  Later, we will
show the relation in terms of components fields.

In deriving the differential equation for $U(t)$, the parameter $t$ is
to be restricted for $|t|<1$ as we will see shortly.  It is
straightforward to obtain the equation
\begin{eqnarray}
\frac{d}{{d} t} U(t) = \frac{1}{2}\Bigl( K_2 + \frac{1}{2}
 U_2(-{\tilde t}) K_4 U_2^{-1}(-{\tilde t})   \Bigr) \frac{U(t)}{1-t^2}.
\end{eqnarray}
The operator $U_2(-{\tilde t}) K_4 U_2^{-1}(-{\tilde t})$ may further be
calculated as
\begin{eqnarray}
U_2(-{\tilde t}) K_4 U_2^{-1}(-{\tilde t})
= \oint \frac{{d}z}{2 \pi i} z(z^4-z^{-4})
\Bigl(\frac{{d}f_2(z;-{\tilde t})}{{d}z}   \Bigr)^2
T\biggl(f_2(z;-{\tilde t})\biggr).
\end{eqnarray}
where the integration path is over the unit circle.  By changing the
variable from $z$ to $u \equiv f_2(z;-{\tilde t})$, we obtain the
expression of the operator 
\begin{eqnarray}
U_2(-{\tilde t}) K_4 U_2^{-1}(-{\tilde t})
= \oint \frac{{d}u}{2 \pi i} u 
\frac{ (1+{\tilde t}^2)(u^4-u^{-4}) + 4 {\tilde t}(u^2-u^{-2})
}{(1+{\tilde t}^2 u^2)(1+{\tilde t}^2 u^{-2})}T(u) .
\label{eq:UK4U}
\end{eqnarray}
It is easy to understand the integration path for $u$ is again over the
unit circle, when the parameter $t$ is real and $|t|<1$.  When we take
$t = -1$, the integration path is not clearly defined.  The value is to
be reached only in the limiting procedure.  The integrand of
(\ref{eq:UK4U}) may be expanded with respect to ${\tilde t}$ since
$|{\tilde t}|<1$ for $|t|<1$.  We finally obtain the differential
equation for $U(t)$,
\begin{eqnarray}
\frac{ d}{{ d} t} U(t) &=& K(t) U(t),\nonumber\\
K(t) &\equiv& \frac{(7-5t)(1+t)}{(1-t)(3-t)^3}K_2 
+ \frac{16(1-t^2)}{(t-3)^4} \sum_2^{\infty}\Bigl(\frac{1-3t}{3-t}\Bigr)^{n-2}K_{2n}.
\end{eqnarray}

Now, we show that $U(t)^{-1}$ has a well-defined normal
ordered expression: 
\begin{eqnarray}
 U(t)^{-1}=
\exp\left(\sum_{n=1}^\infty v_{-2n}L_{-2n}\right)
\exp\left(v_0 L_0\right)
\exp\left(\sum_{n=1}^\infty v_{2n} L_{2n}\right)
\label{eq normal ordered}
\end{eqnarray}
where the coefficients $v_n=v_n(t)$ are finite even in the limit of $t
\rightarrow -1$.  The normal ordered form in eq.~(\ref{eq normal ordered})
are expressed with even modes Virasoro operators since $U_2(t)$ and
$U_{2l}(t)$ themselves are written with $L_{\pm 2}$ and $L_{\pm 2l}$ and the
algebra of even mode operators is closed.

In the following, we take the operator $U(t)$ relating the solutions
$l=1$ and 2 as an example and confirm our claim.  We have not noticed
any difficulty to extend our analysis to other cases.

Taking $\ket{h}$, a highest weight state with the dimension $h$, we
calculate various matrix elements of $U(t)^{-1}$ in terms
of $v_n(t)$ :
\begin{eqnarray}
\bh \Ut \kh&=& e^{v_0 h}, \\
\bh \Ut\,L_{-2}\kh &=& 4h\,v_2\,e^{v_0 h},\\
\bh L_2\,\Ut \kh&=& 4h\,v_{-2}\,e^{v_0 h},\\
\bh L_4\,\Ut \kh&=& \left\{12h\left(v_{-2}\right)^2
  +8h\,v_{-4}\right\}e^{v_0 h},\\
\bh \Ut\,(L_{-2})^2 \kh&=&
 \left\{16h\left(v_2\right)^2+24h\,v_4\right\}e^{v_0 h},
\end{eqnarray}
and so on. Then, if all of these matrix elements are obtained,
we can determine the coefficients $v_n(t)$ iteratively.
In appendix C, we calculate some of the matrix elements using the
definition of $U(t)^{-1}$. As a result, the coefficients
$v_n(t)$ are found to be 
\begin{eqnarray}
 v_0(t)&=&\fron{4}\ln\ckakko{\frac{64(1-t)^3(3-t)^2}{(9t^2-14t+9)^3}},\\
 v_2(t)&=&\frac{(6t^2+4t-18)(3t-1)}{4(9t^2-14t+9)(3-t)},\\
 v_{-2}(t)&=&\frac{3t-1}{2}\kakko{\frac{1-t}{9t^2-14t+9}}^\half,\\
 v_4(t)&=&\frac{8t(1-t)^2(9t^3-4t^2-11t+18)}{(t-3)^2(9t^2-14t+9)^2},\\
 v_{-4}(t)&=&\frac{t}{4}.
\end{eqnarray}
We should emphasize that there is no singularity in $v_n(t)$ in the
limit $t\rightarrow -1$.
Finally, taking the limit, we can obtain the normal ordered expression
of $U(-1)^{-1}$:
\begin{eqnarray}
\label{Eq:normalorder}
&&
U(-1)^{-1}= \lim_{t\rightarrow -1}\Ut\nn
&=&\exp\left(-\frac{1}{2}L_{-2}-\frac{1}{4}L_{-4}+\cdots\right)
\exp\left(-\frac{1}{2}\log 2\,L_0\right)
\exp\left(\frac{1}{8} L_2+\frac{1}{32}L_4 +\cdots\right).
\end{eqnarray}

Now, let us consider the string field transformation
$|\Phi^{(2)}\rangle=U(-1)^{-1} |\Phi^{(1)}\rangle$, by which two theories expanded
around $l=1$ and $l=2$ solutions can be related.  Since the operator
$U(-1)^{-1}$ has the normal ordered expression (\ref{Eq:normalorder}),
the string field transformation has a well-defined Fock space expression,
namely we can obtain transformations for all component fields without
any divergence. 

Write the string field up to level two as
\begin{eqnarray}
\label{Eq:strfield}
 |\Phi^{(1)} \rangle&=& \phi(x)\,c_1\ket{0} \nn
 && +A_\mu(x)\,c_1\alpha^\mu_{-1}\ket{0}+iB(x)\,c_0\ket{0}\nn
 &&+\psi_{\mu\nu}(x)\,c_1\alpha_{-1}^\mu\alpha_{-1}^\nu\ket{0}
  +ia_\mu(x)\,c_1\alpha_{-2}^\mu\ket{0}\nn
 &&+s(x)\,c_{-1}\ket{0}+t(x)\,c_0c_1b_{-2}\ket{0}
+i u_\mu(x)\,c_0 \alpha_{-1}^\mu\ket{0}+\cdots.
\end{eqnarray}
Acting the normal ordered expression (\ref{Eq:normalorder}) of
$U(-1)^{-1}$ on the string field (\ref{Eq:strfield}),\footnote{We have
used the commutation relations
$[L_m,\,\alpha_n^\mu]=-n\alpha_{m+n}^\mu$,
$[L_m,\,c_n]=-(2m+n)c_{m+n}$, 
$[L_m,\,b_n]=(m-n)b_{m+n}$, and
$[L_m,\,\varphi(x)]=-i\sqrt{2\alpha'} \partial_\mu
\varphi(x)\alpha_m^\mu\ \ \ (m\neq 0)$
for any component fields $\varphi(x)$.}
we can easily find transformations for these component fields:
\begin{eqnarray*}
 \phi'(x)&=& 
 \sqrt{2}e^{\frac{\alpha'}{2}\log 2\cdot \partial^2}
\left(\phi(x)+\frac{1}{8}\psi_\mu^\mu(x)+
\frac{\sqrt{2\alpha'}}{4}\partial_\mu a^\mu(x)
-\frac{3}{8} s(x)+\frac{1}{2}t(x)\cdots\right), \\
A'_\mu(x)&=&e^{\frac{\alpha'}{2}\log 2\cdot \partial^2}
  \left(A_\mu(x)+\cdots\right), \\
B'(x)&=&e^{\frac{\alpha'}{2}\log 2\cdot \partial^2}
  \left(B(x)+\cdots\right),\\
\psi'_{\mu\nu}(x)&=&
 e^{\frac{\alpha'}{2}\log 2\cdot \partial^2}
 \left\{-\frac{\sqrt{2}}{4}g_{\mu\nu}\,\phi(x)
 +\frac{1}{\sqrt{2}}\psi_{\mu\nu}(x)\right. \\
 &&\left.-\frac{\sqrt{2}}{4}g_{\mu\nu}\left(
 \frac{1}{8}\psi_\rho^\rho(x)+
 \frac{\sqrt{2\alpha'}}{4}\partial_\rho a^\rho(x)
 -\frac{3}{8} s(x)+\frac{1}{2}t(x)\right)
 +\cdots\right\},\\
a'_\mu(x)&=&
e^{\frac{\alpha'}{2}\log 2\cdot \partial^2}
\left\{\sqrt{\alpha'}\partial_\mu\phi(x)
+\frac{1}{\sqrt{2}}a_\mu(x)\right. \\
&&\left.+\sqrt{\alpha'}\partial_\mu\left(
\frac{1}{8}\psi_\rho^\rho(x)+
\frac{\sqrt{2\alpha'}}{4}\partial_\rho a^\rho(x)
-\frac{3}{8} s(x)+\frac{1}{2}t(x)\right)
+\cdots\right\},\\
s'(x)&=&
e^{\frac{\alpha'}{2}\log 2\cdot \partial^2}
\left\{-\frac{3\sqrt{2}}{2}\phi(x)
+\frac{1}{\sqrt{2}}s(x)\right. \\
&&\left.-\frac{3\sqrt{2}}{2}\left(
\frac{1}{8}\psi_\rho^\rho(x)+
\frac{\sqrt{2\alpha'}}{4}\partial_\rho a^\rho(x)
-\frac{3}{8} s(x)+\frac{1}{2}t(x)\right)
+\cdots\right\},\\
t'(x)&=&
e^{\frac{\alpha'}{2}\log 2\cdot \partial^2}
\left\{\sqrt{2}\phi(x)
+\frac{1}{\sqrt{2}}t(x)\right. \\
&&\left.+\sqrt{2}\left(
\frac{1}{8}\psi_\rho^\rho(x)+
\frac{\sqrt{2\alpha'}}{4}\partial_\rho a^\rho(x)
-\frac{3}{8} s(x)+\frac{1}{2}t(x)\right)
+\cdots\right\},\\
u'_\mu(x)&=&
e^{\frac{\alpha'}{2}\log 2\cdot \partial^2}
\left(\frac{1}{\sqrt{2}}u_\mu(x)+\cdots\right),
\end{eqnarray*}
where the abbreviation denotes contributions from the higher level
component fields. 

The string field transformation has a well-defined expression.  It mixes
tensor fields of various ranks; On each component, it is a non-local
transformation due to infinite derivative terms.

\section{Discussion}

In this paper, we addressed the question how the presence of many
analytic classical solutions for SFT could be consistent with the
physical picture of the tachyon condensation.  Our result suggests that
they are related by a particular type of gauge transformations.  In more
concrete terms, we have seen that the BRS charge for the $l$-th classical
solution $(l \ne 1)$ can be transformed down to that for $l=1$.  The
inverse operation has some subtlety as explained in section 4.  The
transformation is generated by operators $K_n~(n={\rm even})$.  We
observed that the symmetry generated by operators $K_n$ are to be
regarded as the ``global'' part of the SFT gauge symmetry.  The
situation is summarized in the following sequence,
\begin{eqnarray}
&{}&\{{\rm The~stringy~gauge~symmetry}: \Psi'=U^{-1}*\Q U+U^{-1}*\Psi*U\}
\nonumber\\
&{}&\supset
\{{\rm Its~global~subset}: \Psi'=U^{-1}*\Psi*U~{\rm with}~\Q U=0\}
\nonumber\\
&{}&\supset
\{{\rm The~symmetry~generated~with~}K_n: \Psi' = \exp(K(v)) \Psi \}
\nonumber\\
&{}&\supset
\{{\rm The~symmetry~generated~with~}K_n~({n={\rm even}})\}.
\nonumber
\end{eqnarray}
In relating TTK solutions, we have utilized the last subset in the above
sequence.  Generically speaking, solutions are to be related by the
gauge symmetry.  So our approach may be too restrictive and that could
be the reason why we encounter the subtlety.

In order to confirm that the operator relating BRS charges is
well-defined, we studied properties of the operator that transforms
$l=2$ BRS charge into $l=1$ charge and found that it has a well-defined
normal ordered expression in terms of the Virasoro generators.

We studied relations between solutions obtained in Refs.~\cite{rf:TT,
rf:KT}.  Another important direction of investigation is to find how
those solutions could be related to solutions obtained in different
approaches, eg, the level truncation
\cite{rf:SZ-tachyon,rf:MT,rf:GR,rf:KS-tachyon}.  

Most of the works on classical solutions for CSFT have been performed in
the Siegel gauge.  However, the universal solution proposed by
\cite{rf:TT} cannot be in the Siegel gauge as explained in the appendix
D.  A transformation generated by $K_n$ cannot bring a universal
solution into the Siegel gauge: we have to consider more general gauge
transformation.

In relation to the VSFT conjecture and the TTK solutions, recently there
appeared an interesting paper \cite{rf:DrukkerOkawa}.  The VSFT
conjecture on the tachyon vacuum implies that the action
expanded around a TTK solution must be related to VSFT via an
appropriate transformation of the string field.  The authors of
\cite{rf:DrukkerOkawa} discussed this possibility and constructed, with
the level truncation technique and a regulated butterfly state, a
classical solution that could clarify this point.

Before closing, let us add a few remarks.  1) The cohomology analysis
around a classical solution has shown that the ghost numbers of
non-trivial states depend on the value of $l$
\cite{rf:KT,rf:Igarashi4th}.  It would be interesting to see how
operators, eg, $U(t)$ in section 4, relate cohomologically non-trivial
states obtained for various $l$.  That would be another non-trivial test
of those operators and the question certainly deserves further study.
2) We wonder what happens to the symmetries of the SFT defined around
the non-trivial classical solution.  Here we make an observation that
symmetries generated by $K_n$ are broken on these classical solutions.
When we consider an infinitesimal change of the $l$-th solution in
Ref.~\cite{rf:KT}, the function $F_{2l}(w)$ is transformed as shown in
eq.~(\ref{change in F}).  It is easy to see that any choice of the
parameters $\varepsilon_n$ do not leave the function invariant.  This
implies the breaking of the symmetries: the symmetry generated by $K_n$
does not survive the tachyon condensation, at least, in its original
form. 3) The function $F(z)$ in the generic form of classical solution
(\ref{generic form of classical solution}) is to satisfy two
conditions $F(-1/w)=F(w)$ and $F(\pm i)=0$ and is related to another
function $G(z)$ as in eq.~(\ref{rf:F and G}).  Strictly speaking, in
order for (\ref{generic form of classical solution}) to be a
classical solution, these conditions are required to hold only on the
unit circle.  We encounter the same situation in eq.~(\ref{Tl and Tr}).

\vspace{5mm}
\section*{Acknowledgments}
The authors are grateful to I.~Kishimoto, T.~Kugo and T.~Huruya for
discussions.  This work is completed during our stay at the Yukawa
Institute for Theoretical Physics at Kyoto University.  We wish to thank
for their kind hospitality extended to us.  Discussions during the YITP
workshop YITP-W-04-03 on ``Quantum Field Theory 2004'' were useful to
finish this work.  This work is supported in part by the Grants-in-Aid
for Scientific Research No. 13135209 and 15540262 from the Japan Society
for the Promotion of Science.
\appendix

\section{Evaluating $S[g^{-1}*Q_B g, Q_B]$ for $g$ connected to $I$}

We consider the stringy gauge functional $g$ that may be continuously
deformed to the identity $I$: ie, we assume that there exit
a one-parameter family of functionals $g(t)~(0 \le t \le 1)$ so that
$g(0)=I$ and $g(t=1)=g$.  Construct the pure gauge string field as
\begin{eqnarray}
\Psi(t) \equiv g^{-1}(t)*Q_B~g(t).
\label{pure gauge string field}
\end{eqnarray}
By using the properties (\ref{Eq: algebraic structure}), it is easy to
show that the string field $\Psi(t)$ satisfies the equation of motion
(\ref{Eq:eqmotion}).  Note also that $\Psi(t=0)=0$.  Now we may
calculate the variation of the action for $\Psi(t)$ as
\begin{eqnarray}
\frac{d}{dt}S[\Psi(t), Q_B]= \frac{1}{g^2} \int \Bigl(
Q_B \Psi(t)+ \Psi(t)*\Psi(t)
\Bigr) * \frac{d}{dt}\Psi(t)~.
\label{diff of the action}
\end{eqnarray}
The rhs of (\ref{diff of the action}) is proportional to the equation of
motion, and it vanishes. Obviously, it holds that $S[\Psi(t=0),Q_B]=0$.
Therefore $S[\Psi(t),Q_B]=0$ for any value of $t$, in particular $t=1$
\cite{rf:KZ}. 

\section{On the global transformation given in eq.~(\ref{global gauge tf})}

The energy momentum tensor $T(w)$ is expanded by the Virasoro operator
$L_n$ as
\begin{eqnarray}
 T(w)=\sum_{n=-\infty}^\infty L_n w^{-n-1}.
\end{eqnarray}
From commutation relations of $L_n$, we can derive the commutation
relation between $T(w)$ and $T(w')$ as
\begin{eqnarray}
\label{Eq:TT}
 \left[T(w),\,T(w')\right]=-\partial T(w)\delta(w,w')
+T(w)\partial_{w'} \delta(w,w'),
\end{eqnarray}
where the delta function is defined by $\delta(w,w')=\sum_n
w^{-n}{w'}^{n-1}$. 
Here, we define half string operators associated with the
energy-momentum tensor as follows,
\begin{eqnarray}
 {\cal T}_L(f)=\int_{C_{\rm left}}\frac{dw}{2\pi i }f(w)T(w),\ \ \ 
 {\cal T}_R(f)=\int_{C_{\rm right}}\frac{dw}{2\pi i }f(w)T(w).
\end{eqnarray}
Using (\ref{Eq:TT}) and the splitting properties of the delta function
\cite{rf:TT}, we can find the commutation relations between these operators:
\begin{eqnarray}
 \left[{\cal T}_L(f),\,{\cal T}_L(g)\right]
&=&{\cal T}_L((\partial f)g-f\partial g), \\
 \left[{\cal T}_R(f),\,{\cal T}_R(g)\right]
&=&{\cal T}_R((\partial f)g-f\partial g), \\
 \left[{\cal T}_L(f),\,{\cal T}_R(g)\right]&=&0,
\label{Tl Tr}
\end{eqnarray}
where the functions $f(w)$ and $g(w)$ satisfy $f(\pm i)=g(\pm i)=0$. 

If the function $f(w)$ satisfies $f(w)=(dw/d\tilde{w})f(\tilde{w})$ for
$\tilde{w}=-1/w$, we find that
\begin{eqnarray}
\label{Eq:dwT}
 dw f(w) T(w)=d\tilde{w}f(\tilde{w}) \tilde{T}(\tilde{w})
\end{eqnarray}
since $T(w)$ is a primary field with the conformal dimension $2$ for
$c=0$ \cite{rf:RZ}. Using the relation (\ref{Eq:dwT}), we can obtain two properties of the half string
operators for $f(w)$ such that $f(w)=(dw/d\tilde{w})f(\tilde{w})$:
\begin{eqnarray}
\label{Eq:TABTI}
 {\cal T}_R(f)A*B=-A*{\cal T}_L(f)B,\ \ \ 
 {\cal T}_R(f)I+{\cal T}_L(f)I=0,
\label{Tl and Tr}
\end{eqnarray}
where $A$ and $B$ denote arbitrary string fields and $I$ is the identity
string field.  In deriving eq.~(\ref{Eq:TABTI}), it is suffice for the
function $f(w)$ to satisfy the condition, $f(w)=(dw/d{\tilde
w})f({\tilde w})$ for ${\tilde w}=-1/w$, on the unit circle.

We consider gauge transformation with the string functional
\begin{eqnarray}
 g=\exp\left(-{\cal T}_L(f)I \right),
\end{eqnarray}
where $f(w)$ is required to satisfy the condition
$f(w)=(dw/d\tilde{w})f(\tilde{w})$.  It is easy to see that the other
condition $f(\pm i)=0$ follows from the former.  The gauge
transformation is given by
\begin{eqnarray}
 \Psi'=g^{-1}*\Q g+g^{-1}*\Psi*g.
\end{eqnarray}
The first term becomes zero since $[\Q,\,{\cal T}_L(f)]=0$ and $\Q
I=0$. Then, from the equations (\ref{Eq:TABTI}), the gauge
transformation can be rewritten as the string field redefinition
\begin{eqnarray}
 \Psi'= \exp({\cal T}(f))\,\Psi,
\end{eqnarray}
where the operator ${\cal T}(f)$ is defined as ${\cal T}(f)={\cal
T}_L(f)+{\cal T}_R(f)$.

\section{Matrix elements of $U(t)^{-1}$} 

The operators $U_2(\tilde{t})$ and $U_4(t)^{-1}$ can be written using
normal ordered expression as
\begin{eqnarray}
\label{Eq:U2U4normalorder}
U_2(\tilde{t})=
  e^{\frac{\tilde{t}}{2}L_{-2}}\,e^{\frac{1}{2}\log(1-\tilde{t}^2)\,L_0}\,
  e^{-\frac{\tilde{t}}{2}L_2},\ \ \ 
U_4(t)^{-1} =
  e^{\frac{t}{4}L_{-4}}\,e^{\frac{1}{4}\log(1-t^2)\,L_0}\,
  e^{-\frac{t}{4}L_4}.
\end{eqnarray}
Take a normalized highest weight state with the dimension $h$, $\ket{h}$,
and calculate the matrix element $\bra{h}U(t)^{-1}\ket{h}$,
\begin{eqnarray}
\label{Eq:hUh1}
&& \bra{h}U(t)^{-1}\ket{h}= \bra{h}U_4(t)^{-1}U_2(\tilde{t})\ket{h}
= (1-t^2)^{\frac{h}{4}}(1-\tilde{t}^2)^{\frac{h}{2}}
\bra{h}e^{-\frac{t}{4}L_4}\,e^{\frac{\tilde{t}}{2}L_{-2}}\ket{h}~.
\end{eqnarray}

We can derive the recursion relation for matrix elements
$\bra{h}(L_4)^n(L_{-2})^{2n}\ket{h}$,
\begin{eqnarray}
 \bra{h}(L_4)^n(L_{-2})^{2n}\ket{h}=8n(2n-1)(4n+3h-4)
\bra{h}(L_4)^{n-1}(L_{-2})^{2(n-1)}\ket{h},
\end{eqnarray}
which can be solved to give the expression
\begin{eqnarray}
\label{Eq:hL4L2h}
 \bra{h}(L_4)^n(L_{-2})^{2n}\ket{h}=64^n n!
\frac{\Gamma\left(n+\frac{1}{2}\right)}{\Gamma\left(\frac{1}{2}\right)}
\frac{\Gamma\left(n+\frac{3h}{4}\right)}{\Gamma\left(\frac{3h}{4}\right)}.
\label{sol to recursion}
\end{eqnarray}
Using eq.~(\ref{sol to recursion}), we obtain 
\begin{eqnarray}
\label{Eq:heL4eL2h}
 \bra{h}e^{-\frac{t}{4} L_4}\,e^{\frac{\tilde{t}}{2}L_{-2}}\ket{h}
=(1+t\tilde{t}^2)^{-\frac{3h}{4}}.
\end{eqnarray}

Substituting (\ref{Eq:heL4eL2h}) into (\ref{Eq:hUh1}), we reach the 
final expression for $\bra{h}U(-1)^{-1}\ket{h}$:
\begin{eqnarray}
\bra{h}U(t)\ket{h}
= \ckakko{\frac{64(1-t)^3(3-t)^2)}{(9t^2-14t+9)^3}}^\frac{h}{4}.
\end{eqnarray}

Next, let us calculate the matrix element
$\bra{h}U(t)^{-1}L_{-2}\ket{h}$. 
\begin{eqnarray}
\label{Eq:hUL2h}
 \bra{h}U(t)^{-1}L_{-2}\ket{h}
&=& (1-t^2)^{\frac{h}{4}}(1-\tilde{t}^2)^{\frac{h+2}{2}}
\bra{h}e^{-\frac{t}{4} L_4}\,e^{\frac{\tilde{t}}{2}L_{-2}}L_{-2}\ket{h}\nn
&&-2\,h\, \tilde{t}
(1-t^2)^{\frac{h}{4}}(1-\tilde{t}^2)^{\frac{h}{2}}
\bra{h}e^{-\frac{t}{4} L_4}\,e^{\frac{\tilde{t}}{2}L_{-2}}\ket{h}.
\end{eqnarray}
Differentiating eq.~(\ref{Eq:heL4eL2h}) with respect to $\tilde{t}$, we find
\begin{eqnarray}
\label{Eq:heL4eL2L2h}
 \bra{h}e^{-\frac{t}{4} L_4}\,e^{\frac{\tilde{t}}{2}L_{-2}}L_{-2}\ket{h}
=-3h t\tilde{t}(1+t\tilde{t}^2)^{-\frac{3h}{4}-1}.
\end{eqnarray}
Combining the results (\ref{Eq:heL4eL2h}) and (\ref{Eq:heL4eL2L2h})
with (\ref{Eq:hUL2h}),
we find
\begin{eqnarray}
 \bh\Ut L_{-2}\kh
 &=& h\,\frac{\tilde{t}(t \tilde{t}^2-3t-2)}{1+t\tilde{t}^2}\,
\left\{
\frac{(1-t^2)(1-\tilde{t}^2)^2}{(1+t\tilde{t}^2)^3}\right\}^{
\frac{h}{4}}\nn
&=&h\,\frac{2(3t^2+2t-9)(3t-1)}{(9t^2-14t+9)(3-t)}\,
\ckakko{\frac{64(1-t)^3(3-t)^2)}{(9t^2-14t+9)^3}}^\frac{h}{4}.
\end{eqnarray}

Other matrix elements may be calculated in a similar manner. Using
the normal ordered expression of $U_2({\tilde t})$ and $U_4(t)$, we easily find 
\begin{eqnarray}
 \bra{h}L_2 U(t)^{-1}\ket{h}
&=&(1-t^2)^{\frac{h+2}{4}}(1-\tilde{t}^2)^{\frac{h}{2}}\nn
&&\hspace*{-1.5cm}\times \sum_{n=0}^\infty \frac{1}{(2n+1)!\,n!}
\left(\frac{\tilde{t}}{2}\right)^{2n+1}
\left(-\frac{t}{4}\right)^n
\bra{h}L_2\,(L_4)^n (L_{-2})^{2n+1}\ket{h},
\label{first eq}\\
\bra{h}L_4U(t)^{-1}\ket{h}
&=& (1-t^2)^{\frac{h+4}{4}}(1-\tilde{t}^2)^\frac{h}{2}
\bra{h}L_4\,e^{-\frac{t}{4}L_4}\,e^{\frac{\tilde{t}}{2}L_{-2}}\ket{h}\nn
&&+(1-t^2)^{\frac{h}{4}}(1-\tilde{t}^2)^\frac{h}{2}\,2h\,t\,
\bra{h}e^{-\frac{t}{4}L_4}\,e^{\frac{\tilde{t}}{2}L_{-2}}\ket{h},
\label{second eq}\\
\bra{h}U(t)^{-1}(L_{-2})^2\ket{h}&=&
(1-t^2)^{\frac{h}{4}}(1-\tilde{t}^2)^\frac{h+4}{2}
\bra{h}e^{-\frac{t}{4}L_4}\,e^{\frac{\tilde{t}}{2}L_{-2}}(L_{-2})^2\ket{h}\nn
&&-4(h+1)(1-t^2)^{\frac{h}{4}}\tilde{t}\,(1-\tilde{t}^2)^\frac{h+2}{2}
\bra{h}e^{-\frac{t}{4}L_4}\,e^{\frac{\tilde{t}}{2}L_{-2}}\,L_{-2}\ket{h}\nn
&&+4h(h+1)(1-t^2)^{\frac{h}{4}}\tilde{t}^2
(1-\tilde{t}^2)^\frac{h}{2}
\bra{h}e^{-\frac{t}{4}L_4}\,e^{\frac{\tilde{t}}{2}L_{-2}}\ket{h}.
\label{third eq}
\end{eqnarray}
We can calculate (\ref{first eq}) by using
\begin{eqnarray}
 \bra{h}L_2\,(L_4)^n (L_{-2})^{2n+1}\ket{h}=
4^{3n+1}h\,n!\,
\frac{\Gamma\left(n+\frac{3}{2}\right)}{\Gamma\left(\frac{3}{2}\right)}
\frac{\Gamma\left(n+\frac{3h}{4}+\frac{1}{2}\right)}{
\Gamma\left(\frac{3h}{4}+\frac{1}{2}\right)},
\end{eqnarray}
and eqs.~(\ref{second eq}) and (\ref{third eq}) can be evaluated by
using eqs.~(\ref{Eq:heL4eL2h}),
(\ref{Eq:heL4eL2L2h}) and
\begin{eqnarray}
 \bra{h}L_4\,e^{-\frac{t}{4}L_4}e^{\frac{\tilde{t}}{2}L_{-2}}\ket{h}
&=&3h\,\tilde{t}^2\,(1+t\tilde{t}^2)^{-\frac{3h}{4}-1},\nonumber\\
 \bra{h}e^{-\frac{t}{4}L_4}e^{\frac{\tilde{t}}{2}L_{-2}}
(L_{-2})^2\ket{h}
&=&3h\,t\,(1+t\tilde{t}^2)^{-\frac{3h}{4}-2}(
-2-2\,t\,\tilde{t}^2+3h\,t\,\tilde{t}^2)\nonumber.
\end{eqnarray}
Finally, we obtain the expressions for the matrix elements,
\begin{eqnarray}
 \bh L_2\,\Ut\kh
  &=&2h\,\tilde{t}\left(\frac{1-t^2}{1+t\tilde{t}^2}\right)^\frac{1}{2}\,
     \left\{\frac{(1-t^2)(1-\tilde{t}^2)^2}{(1+t\tilde{t}^2)^3}\right\}^{
\frac{h}{4}}\nn
  &=&2h\,\frac{(3t-1)(1-t)^\half}{(9t^2-14t+9)^\half}\,
\ckakko{\frac{64(1-t)^3(3-t)^2)}{(9t^2-14t+9)^3}}^\frac{h}{4},
\nonumber\\
 \bh L_4\,\Ut\kh
   &=& h\,\frac{2t+3\tilde{t}^2-t^2\tilde{t}^2}{1+t\tilde{t}^2}\,
      \left\{\frac{(1-t^2)(1-\tilde{t}^2)^2}{(1+t\tilde{t}^2)^3}\right\}^{
\frac{h}{4}}\nn
   &=&h\,\frac{-9t^3+17t^2-3t+3}{9t^2-14t+9}\,
\ckakko{\frac{64(1-t)^3(3-t)^2)}{(9t^2-14t+9)^3}}^\frac{h}{4},
\nonumber\\
 \bh\Ut\,(L_{-2})^2\kh
  &=& \left[16h(h+1)\left\{\frac{\tilde{t}(t\tilde{t}^2-3t-2)}{
         4(1+t\tilde{t}^2)}\right\}^2\right.\nn
  &&\left.-24h\,\frac{t(1-\tilde{t}^2)(2+t\tilde{t}^2)}{8(1+t\tilde{t}^2)^2}\,
       \right]\,\left\{\frac{(1-t^2)(1-\tilde{t}^2)^2}{(1+t\tilde{t}^2)^3}
         \right\}^{\frac{h}{4}}\nn
  &=&\left[16h(h+1)\left\{
    \frac{(3t-1)(3t^2+2t-9)}{2(3-t)(9t^2-14t+9)}\right\}^2
   \right.\nn
    && \hspace{-20mm}\left.-24h\,\frac{8t(1-t)^2(9t^3-4t^2-11t+18)}{
       (t-3)^2(9t^2-14t+9)^2}\right]\,
\ckakko{\frac{64(1-t)^3(3-t)^2)}{(9t^2-14t+9)^3}}^\frac{h}{4}.
\end{eqnarray}

\section{The universal solutions are not in the Siegel gauge}

In this appendix, we show that the universal solution given in
Ref.~\cite{rf:TT} 
does not satisfy the Siegel gauge condition.

The universal solution is written in the following form
\begin{eqnarray}
 \ket{\Psi_0}&=&Q_\rL(F)\ket{I}+\CL(G)\ket{I},
 \label{Eq:Psi0}\\
 &&G=-\frac{(\partial F)^2}{1+F}.
\label{G by F}
\end{eqnarray}

Let us search for the function $F(w)$ which satisfies the Siegel gauge condition
\begin{eqnarray}
 \label{Eq:b0psi}
0= b_0\ket{\Psi_0}
  =\int_\Cl\dz{w}F(w)b_0\JB(w)\ket{I}
  +\int_\Cl\dz{w}G(w)b_0c(w)\ket{I},
\label{Siegel gauge on Usol}
\end{eqnarray}
relying on the conformal technique.  First note that the identity state
$\ket{I}$ can be written as $\ket{I} =U^{-1}_{I\circ h\circ I}\ket{0}$.
That is, the state can be expressed with the operator for the conformal
transformation,
\begin{eqnarray*}
 I\circ h\circ I=\frac{w^2-1}{2w} \equiv g(w).
\end{eqnarray*}

Using $Q_n\ket{0}=0\,(n\geq0)$ and 
\begin{eqnarray*}
 U_g\JB(w)U_g^{-1}
  =[\partial g(w)]^{+1}\JB(g(w))
  =\partial g(w)\sum^{\infty}_{n=-\infty}(g(w))^{-n-1}Q_n,
\end{eqnarray*}
we find 
\begin{eqnarray*}
 \JB(w)\ket{I}
  =U^{-1}_g\sum^\infty_{n=1}\partial g(w)(g(w))^{n-1}Q_{-n}\ket{0}.
\end{eqnarray*}
So we obtain the expression for the first term of eq.~(\ref{Eq:Psi0})
\begin{eqnarray}
 \label{Eq:(1)}
 Q_\rL(F)\ket{I}&=&U^{-1}_g\sum^\infty_{n=1}\alpha_nQ_{-n}\ket{0}
\end{eqnarray}
where $\alpha_n$ is given as
\begin{eqnarray}
 \alpha_n=\int_\Cl\dz{w}F(w)(g(w))^{n-1}\partial g(w).
\label{Eq:alpha n}
\end{eqnarray}

We rewrite the second term $C_\rL(G)\ket{I}$ in a similar manner: since
\begin{eqnarray*}
 U_gc(w)U^{-1}_g
  =[\partial g(w)]^{-1}c(g(w))
  =[\partial g(w)]^{-1}\sum^\infty_{n=-\infty}(g(w))^{-n+1}c_n, 
\end{eqnarray*}
we obtain
\begin{eqnarray}
 \label{Eq:(2)}
  \CL(G)\ket{I}=U^{-1}_g\sum^\infty_{n=-1}\beta_nc_{-n}\ket{0},
\end{eqnarray}
with $\beta_n$ given as
\begin{eqnarray}
 \beta_n \equiv \int_\Cl\dz{w}G(w)(g(w))^{n+1}(\partial g(w))^{-1}.
\end{eqnarray}

From eqs.~(\ref{Eq:(1)}) and (\ref{Eq:(2)}), the universal
solution is now rewritten as
\begin{eqnarray}
 \ket{\Psi_0}
  =U^{-1}_g\ckakko{\sum^\infty_{n=1}\alpha_nQ_{-n}\ket{0}
  +\sum^\infty_{n=-1}\beta_nc_{-n}\ket{0}}.
\label{eq univ sol}
\end{eqnarray}

Let us write the gauge condition for the universal solution
written as eq.~(\ref{eq univ sol}).  First note 
\begin{eqnarray}
 U_gb_0U_g^{-1}
  =\oint\dz{w}wU_gb(w)U_g^{-1}
  =\oint\dz{w}w(\partial g(w))^2b(g(w))
  =\sum^{\infty}_{n=-\infty}\gamma_nb_{-n},
\end{eqnarray}
where $\gamma_n$ is
\begin{eqnarray}
 \gamma_n
  =\oint\dz{w}w(\partial g(w))^2(g(w))^{n-2}
  =\oint\dz{w}w\kakko{\frac{1+w^2}{2w^2}}^2\kakko{\frac{w^2-1}{2w}}^{n-2}.
\end{eqnarray}
After some calculation, we find that $\gamma_n$ vanish for $n \leq -1$
and $n=$odd.  Therefore the condition may be written 
\begin{eqnarray*}
 0=b_0\ket{\Psi_0}=U_g^{-1}\ckakko{
  \sum^\infty_{m=0}\sum^\infty_{n=1}
  \gamma_{2m}\alpha_n\cdot b_{-2m}Q_{-n}\ket{0}
  +\sum^\infty_{m=0}\sum^\infty_{n=-1}
  \gamma_{2m}\beta_n\cdot b_{-2m}c_{-n}\ket{0}},
\end{eqnarray*}
or,
\begin{eqnarray}
 \label{Eq:(3)}
 \sum^\infty_{m=0}\sum^\infty_{n=1}
  \gamma_{2m}\alpha_n\cdot b_{-2m}Q_{-n}\ket{0}
  +\sum^\infty_{m=0}\sum^\infty_{n=-1}
  \gamma_{2m}\beta_n\cdot b_{-2m}c_{-n}\ket{0}=0.
\end{eqnarray}
Explicitly writing the state $b_{-2m}Q_{-n}\ket{0}(m\geq1,n\geq1)$ as
\begin{eqnarray*}
 b_{-2m}Q_{-n}\ket{0}=b_{-2m}c_0L^X_{-n}\ket{0}+\cdots,
\end{eqnarray*}
we realize that the first and second terms in eq.~(\ref{Eq:(3)}) cannot
cancel with each other.  Thus $\alpha_n=0$ \ ($n\geq1$) as well as $\beta_n=0$
\ ($n\geq-1$).  Rewriting (\ref{Eq:alpha n}) with $w=e^{i\sigma}$, we find 
\begin{eqnarray*}
 \alpha_n
  =i^{n-1}\int^{\frac{\pi}{2}}_{-\frac{\pi}{2}}\frac{d\sigma}{2\pi}
  F(\sigma)\kakko{\sin\sigma}^{n-1}\cos\sigma
=\int^1_{-1}\frac{dx}{2\pi}\tilde{F}(x)x^{n-1}
\end{eqnarray*}
In the last expression, we changed the variable as $x=\sin\sigma$ and
used the notation $F(\sigma)=\tilde{F}(x)$.  Clearly, the Siegel
gauge condition requires the vanishing of the function,
$\tilde{F}(x)=0$, therefore $\ket{\Psi_0}=0$.

In conclusion, the universal functions cannot be the Siegel gauge.

\end{document}